\documentclass[12pt]{article}
\usepackage{graphicx,amssymb,amsmath}

\def\beq{\begin{equation}}
\def\eeq{\end{equation}}
\def\ba{\begin{eqnarray}}
\def\ea{\end{eqnarray}}
\def\ar{\begin{array}}
\def\ear{\end{array}}
\def\pa{\partial}

\def\nn{\nonumber\\}

\def\ga{\gamma}

\def\a{\alpha}
\def\b{\beta}

\def\eps{\epsilon}
\def\vp{\varphi}
\def\ve{\varepsilon}

\def\ri{{\rm i}}
\def\GeV{{\rm GeV}}
\def\eV{{\rm eV}}

\def\a{\alpha}
\def\da{{\dot\alpha}}
\def\b{\beta}
\def\db{{\dot\beta}}
\def\g{\gamma}

\def\ve{\varepsilon}
\def\vp{\varphi}

\def\bn{{\bar\nu}}
\def\bN{{\bar{N}}}

\def\ta{\tilde{a}}

\def\MM{{\mathbb M}}

\def\tF{\widetilde{F}}
\def\tG{\widetilde{G}}
\def\tW{\widetilde{W}}

\def\X{{\mathfrak{X}}}

\def\ri{{\rm i}}

\def\cJ{{\cal J}}
\def\cO{{\cal O}}
\def\cC{{\cal C}}
\def\cD{{\cal D}}

\def\cQ{{\cal Q}}

\def\mL{{\mathfrak{L}}}

\def\sp{\not{\!p}}

\def\spa{\not{\!\partial}}

\begin{document}
\begin{center}
{\bf \Large Axions without Peccei-Quinn Symmetry}\\[1.5cm]

{\bf Adam Latosinski$^1$, Krzysztof A. Meissner$^{1,2}$ and Hermann Nicolai$^3$}

\vspace{1cm}
{{\it $^1$ Faculty of Physics,
University of Warsaw,
Ho\.za 69, Warsaw, Poland\\
$^2$ Soltan Institute for Nuclear Studies, \'Swierk, Poland\\
$^3$ Max-Planck-Institut f\"ur Gravitationsphysik
(Albert-Einstein-Institut)\\
M\"uhlenberg 1, D-14476 Potsdam, Germany\\
}}
\end{center}

\vspace{2cm}

{\footnotesize
{We argue that the axion arising in the solution
of the strong CP problem can be identified with the Majoron, the
(pseudo-)Goldstone boson of spontaneously broken lepton number
symmetry. At low energies, the associated $U(1)_L$ becomes, via
electroweak parity violation and neutrino mediation, indistinguishable
from an axial Peccei-Quinn symmetry in relation to the strong
interactions. The axionic couplings are then {\em fully computable}
in terms of known SM parameters and the (as yet unknown) Majorana mass
scales, as we illustrate by computing the effective couplings to photons
and quarks at two loops. Together with previous results our proposal
provides further evidence that the known particle physics phenomena
can all be explained without introducing intermediate scales of any kind
between the electroweak scale and the Planck scale.}
}

\newpage
\section{Introduction}

The solution of the strong CP problem by means of the Peccei-Quinn
mechanism \cite{PQ} is commonly assumed to require the presence of
a {\em chiral} $U(1)_{PQ}$ symmetry (Peccei-Quinn symmetry) which is not
part of the standard model (SM), as well as an independent new scale
$ \geq \! \cO (10^{10})\,\GeV$ beyond the SM. When spontaneously
broken, the PQ symmetry
gives rise to a (pseudo-)Goldstone boson, the {\em axion} \cite{W1,W2}.
The latter is usually described by a pseudoscalar field transforming
by constant shifts under $U(1)_{PQ}$. The absence of CP violation in the
strong interactions is then explained by the fact that any contribution
to the $\theta$ parameter can be absorbed into such a shift, so the
problem is solved if the axion vacuum expectation value dynamically adjusts
itself to zero \cite{VW}. To accommodate the extra $U(1)_{PQ}$ the
available models realizing this idea invariably need to introduce (so far
unobserved) new particles and large scales beyond the SM, such as new heavy
quarks or non-standard Higgs fields \cite{Kim,Dine}.

In \cite{MN} a minimal extension of the SM was proposed, based on
the hypothesis that quantum mechanically broken conformal symmetry
stabilizes the electroweak hierarchy, with only the right-chiral
neutrinos $\nu_R^i$ and one complex scalar field
\beq\label{phi}
\phi(x) = \vp (x) \exp \left(\frac{ia(x)}{\sqrt{2}\mu}\right)
\eeq
as new ingredients (for alternative models based on conformal
symmetry, see \cite{CW}; a similarly minimalistic scenario without exact
conformal symmetry had already been developed in great detail in \cite{Shap}).
The field $\phi$ is a singlet under the SM gauge symmetries and couples only to
right-chiral neutrinos, see (\ref{L}) below. If $\phi$ acquires a vacuum expectation value by
(possibly radiatively induced) spontaneous symmetry breaking, a Majorana
mass term is generated for the right-chiral neutrinos. The phase $a(x)$
then gives rise to a (pseudo-)Goldstone particle (called `Majoron')
associated with the spontaneous breaking of global $U(1)_L$ lepton
number symmetry \cite{Mo1}. The crucial feature of our proposal is that
it requires all mass scales to arise from the quantum mechanical breaking
of classical conformal invariance. Therefore in any consistent implementation
of this scheme {\em there cannot exist intermediate scales of any kind
between the electroweak scale and the Planck scale.} This holds
in particular true for the masses of the light neutrinos whose smallness
is naturally explained here with appropriate neutrino Yukawa couplings
$\sim \cO(10^{-5})$~\footnote{Recall that the appearance of a similar ratio for
 the charged leptons is an experimental fact: $m_e/m_\tau < 10^{-5}$.}
 and without the need to introduce a large
Majorana mass `by hand'. In this paper we show that likewise,
and contrary to widely held expectations, no extra large scale is
required for the solution of the strong CP problem either.

As argued in \cite{MN} the Majoron has several features in common
with the axion, and the smallness of its couplings can be tied to
the smallness of neutrino masses. In this Letter, we go one step further
and propose that the Majoron actually {\em is} the axion, with
{\em computable} effective couplings to SM particles, and the
neutrino Yukawa couplings as the only unknown parameters (a possible
link between light neutrinos and the invisible axion had already
been suggested in \cite{Mo3,LPY}). In other words, we claim
that lepton number symmetry $U(1)_L$ is transmuted, via electroweak
parity violation and neutrino mixing, into a $U(1)$ symmetry that,
in relation to the strong interactions, is indistinguishable from
the standard axial Peccei-Quinn symmetry at low energies.
We present exact expressions for the (UV finite)
two-loop integrals describing the coupling of the axion to photons
and (light) quarks; the main technical novelty here is the
consistent use of the off-diagonal neutrino propagators
(\ref{nuProp}) below. From the quark couplings one can estimate the
coupling of the axion to gluons, which comes out naturally tiny.

On general grounds the effective couplings of $a(x)$ can only be
of a very restricted type. Because Goldstone bosons interact only
via derivatives, the perturbative
effective action at low energies contains only terms $\propto\X^\mu
\partial_\mu a$, where $\X^\mu$ are local expressions in the SM
quantum fields. At lowest order there are only three
candidates for $\X^\mu$: ($i$) a Chern-Simons current, which by
partial integration is equivalent to a  coupling $a \, {\rm Tr}\,
W_{\mu\nu}\tW^{\mu\nu}$ (where $W_\mu$ can be any SM gauge
connection), ($ii$) a vector current $\cJ_V^\mu$ and ($iii$) an axial
current $\cJ_A^\mu$. Being mediated by the weak interactions the
fermionic bilinears contributing to $\X^\mu$ and involving charged
SM fermions all appear in `V\ --\ A' form. Therefore, whenever
$\partial_\mu\cJ_V^\mu\approx 0$ by some approximate\footnote{By
  this we mean neglecting all terms involving neutrinos or the scalar field
  $\phi$ in the relevant currents, as well as baryon or lepton number violating
  `sphaleron-like' contributions, because these will give negligible
  contributions to all processes considered in this paper.}
 conservation law, $a(x)$ couples like a {\em pseudoscalar} to
photons, gluons, quarks  and electrons.

\section{Neutrino Lagrangian and propagators}

We refer to \cite{EPP,Pok} for basic properties of the SM, and here only quote the
Yukawa couplings
\ba\label{L}
\mL_{\rm Y}\!\!\!&=&\!\!\!
\Big(\overline{L}^i\Phi Y_{ij}^E E^j  +
\overline{Q}^i\Phi Y_{ij}^D D^j+\overline{Q}^i\ve\Phi^* Y_{ij}^U U^j\nn
&& + \, \overline{L}^i\ve\Phi^\ast Y_{ij}^\nu N^j+\phi N^{i T}
\cC^{-1}Y^M_{ij} N^j+{\rm h.c.}\Big)
\ea
and the neutrino terms in the Lagrangian, see (\ref{Lneutrino}) below. Here $Q^i$
and $L^i$ are the left-chiral quark and lepton doublets, $U^i$ and $D^i$ the
right-chiral up- and down-like quarks, while $E^i$ are the right-chiral
electron-like leptons, and $N^i\equiv\nu^i_R$ the right-chiral neutrinos (we
suppress all indices except the family indices $i,j=1,2,3$). $\Phi$ is the usual
Higgs doublet, and $\phi$ is the new complex scalar field introduced in (\ref{phi}).
As is well known, one can use global redefinitions of the fermion
fields to transform the Yukawa matrices $Y_{ij}^E$, $Y_{ij}^U$ and $Y_{ij}^M$ to
real diagonal matrices. By contrast, the matrices $Y_{ij}^D$ and $Y^\nu_{ij}$ may
exhibit (strong) mixing. Besides the standard (local) $SU(3)_c\times SU(2)_w \times
U(1)_Y$ symmetries, the Lagrangian (\ref{L}) admits two {\em global} $U(1)$
symmetries, baryon number symmetry $U(1)_B$ and lepton number symmetry $U(1)_L$. The
latter is associated with the Noether current
\beq\label{JL}
\cJ^\mu_L := \overline{L}^i\ga^\mu L^i
+\overline{E}^i\ga^\mu E^i+\overline{N}^i\ga^\mu N^i
-2 i \phi^\dagger\! \stackrel{\leftrightarrow}{\pa^\mu} \!\phi
\eeq
The fact that $\phi$ carries lepton charge is crucial for the proposed transmutation
of $U(1)_L$ into a PQ-like symmetry.

For the computation of loop diagrams it is convenient to employ
$SL(2,\mathbb{C})$ spinors \cite{MN}. With $\nu^i_L \equiv
\frac12(1-\g^5)\nu^i\equiv \bn^{i\da}$ and $\nu^i_R \equiv\frac12
(1+\ga^5)\nu^i \equiv N^i_\a$, the neutrino part of the free
Lagrangian reads (see \cite{BW} for conventions)
\beq\label{Lneutrino}
\mL = \frac{\ri}2 \Big(
  \nu^{i\a} \!\spa_{\a\db} \bn^{i\db}
      + N^{i\a} \!\!\spa_{\a\db} \bN^{i\db} \Big)  \,
      + \, m_{ij}\, \nu^{i\a} N^j_{\a} +
  \frac12 M_{ij} \, N^{i\a} N^j_{\a} \, + \; c.c.
\eeq
after spontaneous breaking of conformal and electroweak symmetries. Consequently,
the (complex) Dirac and Majorana mass matrices are given by $m_{ij} =
Y^\nu_{ij}\langle H \rangle$ and $M_{ij}= Y^M_{ij} \langle \vp \rangle$,
respectively (where $\langle H\rangle^2\equiv \langle \Phi^\dagger\Phi\rangle$).
Rather than diagonalize the fields w.r.t. these mass terms, we work with {\em
non-diagonal propagators} and the interaction vertices from (\ref{L}). Defining
\ba
\cD(p) &:=& \Big[ p^4 - p^2(M^{-1}m^Tm^*M + m^\dag m + M^*M) \nn
&&\quad\qquad     + \, m^\dag m M^{-1} m^Tm^*M \Big]^{-1}
\ea
we obtain the matrix propagators (in momentum space)
\ba\label{nuProp}
\langle \nu^i_\a \nu^j_\b \rangle &=& \ri\left[m^*M \cD(p)m^\dagger\right]^{ij}
           \ve_{\a\b}  \nn
\langle \nu^i_\a \bar\nu^j_{\db} \rangle &=& \ri\big[(m^T)^{-1}
     \big\{p^2- MM^* -
(M^*)^{-1}m^\dagger mM^* \big\}\cD(p)^* m^T\big]^{ij}
\sp_{\a\dot\b} \nn
\langle N^i_\a N^j_\b \rangle &=& \ri\left[M^* p^2\cD(p)^*\right]^{ij}
         \ve_{\a\b}  \nn
\langle N^i_\a \bar{N}^j_{\dot\b} \rangle &=&
\ri\left[\left(p^2-M^{-1}m^Tm^*M\right)\cD(p)\right]^{ij} \sp_{\a\db} \nn
\langle \nu^i_\a N^j_\b \rangle &=&
\ri\left[m^*\big\{p^2- (M^*)^{-1}m^\dag mM^* \big\}\cD(p)^*\right]^{ij}
         \ve_{\a\b}  \nn
\langle \nu^i_\a \bar{N}^j_{\dot\b} \rangle &=&
-\ri\left[m^*M \cD(p)\right]^{ij} \sp_{\a\dot\b}\, ,
\ea
together with their complex conjugate components. Evidently,
these propagators allow for maximal mixing in the sense that every
neutrino component can oscillate into any other (also across families).
For the UV finiteness of the diagrams to be computed below it is
essential that some of the propagator components fall off like
$\sim p^{-3}$, unlike the standard Dirac propagator. Taking $M_{ij}$
diagonal it is not difficult to recover the mass eigenvalues as
predicted by the standard see-saw formula \cite{Min,seesaw,Yan,Mo2}.

With the above propagators and the (extended) SM Lagrangian we can now
proceed to compute various effective low energy couplings involving the
`axion' $a$ which are mediated by neutrino mixing via two or three-loop
diagrams. Here we present only the results for photon-axion and
quark-axion couplings, cf. the diagrams depicted below. Further
results and detailed derivations will be given in a forthcoming
publication \cite{LMN}.


\section{Photon-axion vertex}

For the low energy effective action we need only retain contributions where all
particles circulating inside the loops are much heavier than the external particles.
As our first example we determine the effective coupling of the axion to photons via
the two-loop diagram in Fig.~1. For small axion momentum $q^\mu= k_1^\mu - k_2^\mu$
it is possible to derive a closed form expression for the two-loop integral and for
arbitrary mixing matrices \cite{LMN}. Setting $\mu=\langle\vp\rangle$ in (\ref{phi})
and denoting by $M_j$ the eigenvalues of the (diagonal) matrix $M_{ij}$, a lengthy
calculation gives the expected kinematical factor
$\eps^{\mu\nu\lambda\rho}F_{\mu\nu} F_{\lambda\rho}
\propto \eps^{\mu\nu\lambda\rho}k_{1\lambda} k_{2\rho}$
with coefficient function
\ba
&&\!\!\!\!\!\!
\frac{\ri e^2g_w^2}{64\sqrt{2}\pi^4}
\sum_{i,j} \frac{|m_{ij}|^2M_j^2}{\langle\vp\rangle} \,
\! \int_0^1 dx \int_0^1\!\! dy \int_0^1\!\! dz\int_0^1 \!\! dt\,
x(1-x)y^2 z(1-z) t^3 \,\times \\
&& \times\left\{\frac{-yt - 6(1-y)zt}{\MM_{ij}^4(x,y,z,t)} \; +
 \frac{y(1-y)[ -2y - 3(1-y)zt][(2-t) m_{e_i}^2 - t(1-t)^2 k_2^2]}
     {\MM_{ij}^6(x,y,z,t)} \right\}\nn
\ea
where $m_{e_i} \equiv (m_e, m_\mu, m_\tau)$ and
\ba
\MM^2_{ij}(x,y,z,t) :=  xyzt M_j^2 + (1-y)\Big[ zt M_W^2
   -yt(1-t)k_2^2 + y(1-zt) m_{e_i}^2\Big]
\ea
The above integral is cumbersome to evaluate in general form, but for
small photon momenta $k_1^\mu\approx k_2^\mu$ we get
\beq
(7) \; \approx \;
\frac{\ri\alpha_{em} \alpha_w}{72\sqrt{2}\pi^2}
\sum_{i.j} \frac{|m_{ij}|^2}{\langle\vp\rangle M_j^2}
            \left(\log \frac{M_j^2}{m_e^2}\right)^2
\eeq
Of course, the precise value of the effective low energy coupling
depends on the (unknown) values of the Yukawa mass matrices $m_{ij}$
and $M_{ij}= M_j \delta_{ij}$. For $M_j \! \sim \! M\!\sim\!\langle\vp\rangle$
the axion-photon vertex is well approximated by
\beq\label{fa}
\mL_{\rm eff}^{a\ga\ga} =
\frac1{4f_a} a \, F^{\mu\nu} \tF_{\mu\nu}  \, , \;\;
f_a^{-1} =
\frac{\alpha_{em}\alpha_w \sum m_\nu}{72\sqrt{2}\pi^2 M^2}
\left(\log\frac{M^2}{m_e^2}\right)^2
\eeq
with the standard see-saw relation $\sum m_\nu \sim \sum |m_{ij}|^2/M$.
Substituting numbers we find $f_a= \cO( 10^{16} \, \GeV)$ which is
outside the range of existing or planned experiments \cite{OSQAR}.
Thus the smallness of axion couplings gets directly tied to the smallness
of the light neutrino masses via (\ref{fa}).

\section{Quarks and gluons}

The effective low energy couplings to light quarks can be analyzed in a similar way.
With $P_L\equiv \frac12(1-\gamma^5)$ we parametrize these couplings
as\footnote{While we use capital letters
  $U,D,...$ in (\ref{L}) to designate {\em chiral} spinors, we use small letters
  $u,d,...$ for the full (non-chiral) spinors here and below.}
\beq\label{qa}
\mL_{\rm eff}^{aqq} = \ri\partial_\mu a \Big(
c_{ij}^{aUU} \bar{u}^i\gamma^\mu P_L u^j +
c_{ij}^{aDD} \bar{d}^i\gamma^\mu P_L d^j \Big)
\eeq
Again one can obtain an exact formula for the (UV finite) two-loop integrals;
e.g. for the up-like quarks we get
\ba\label{ca}
c^{aUU}_{ij} &=&
\sum_{k,r,s} \frac{g_w^4 |m_{rs}|^2 M_s^2  V^{ik}
(V^\dagger)^{kj}}{128\sqrt{2}\pi^4\langle\vp\rangle} \; \times \\
&& \!\!\!\!\!\!\!\!\!\!
\int_0^1dx \int_0^1dy \int_0^1dt \int_0^1dz \,\, x(1-x)y^3(1-z)t^3
 \; \times\nn
&& \!\!\!\!\!\!\!\!\!\!\!\!\!\!\!\!\!\!\!\!\!\!\!\!\!\!\!\!\!\!
\frac{-1+3y + 3(1-y)zt}{\left[xyzt M_s^2 + (1-y)\{ yt(1-z) M_W^2
        + zt m^2_{e_r} + y(1-t) m^2_{D_k}\}\right]^2} \nonumber
\ea
with the CKM matrix $V^{ij}$. A similar (but not the same) formula
is obtained for $c^{aDD}_{ij}$ \cite{LMN}. In principle, there are
also contributions from diagrams with $Z$-boson exchange, but these
can be disregarded for the effective low energy Lagrangian because they
involve a purely neutrino triangle with one light neutrino (which is
lighter than any external quark). To estimate the integral, we set
$m_{e_i} = m_{d_i}=0$ in (\ref{ca}) (which still leaves a convergent
integral that can be calculated exactly \cite{LMN}). Because the CKM
matrix is unitary, both $c^{aUU}_{ij}$ and $c^{aDD}_{ij}$ then become
proportional to $\delta_{ij}$ to leading order. Keeping only one lepton
flavor in (\ref{ca}) we here quote the result only for two limiting
cases: for $M_j \!\sim \! M \gg \! M_W$ we get
\beq
c^{aUU}_{ij} = \sum_{k,l}
\frac{\alpha_w^2 |m_{kl}|^2}{128\sqrt{2} \pi^2 \langle\vp\rangle M_j^2}
\left[ \left(\log \frac{M_j^2}{M_W^2} -2\right)^2 + \frac{2\pi^2}3 \right]
   \delta_{ij}
\eeq
If instead $M \!\sim\! M_W$ the exact result replaces the square bracket
by $0.71$. Note that the Majorana mass $M$ is much closer to the weak
scale in \cite{Shap,MN} than in the usual see-saw scenario, favoring
the second value.

By the approximate conservation of the up and down quark
vector currents, we can now drop the vectorlike contribution in the
effective Lagrangian which thus becomes purely axial to leading order,
{\it viz.}
\beq\label{qa1}
\mL_{\rm eff}^{aqq} \, \rightarrow \, \ri\partial_\mu a \Big(
  g_{aUU}^{-1} \bar{u}^j\gamma^5 \gamma^\mu u^j +
  g_{aDD}^{-1} \bar{d}^j\gamma^5 \gamma^\mu d^j \Big)
\eeq
At subleading order off-diagonal contributions to $c_{ij}^{aUU}$ and
$c_{ij}^{aDD}$ will appear with both vector and axial vector
interactions. The numerical values of the effective coupling constants
can be read off from the above results. Their precise values are
subject to the same caveats as mentioned before (\ref{fa}). With the
same assumptions on the Yukawa mass matrices as for (\ref{fa}) we get
\beq\label{qa2}
g_a^{-1}\equiv
g_{aUU}^{-1} \sim g_{aDD}^{-1} \sim
\cO(10^{-3}) \frac{\alpha_w^2 \sum m_\nu}{M^2}
\eeq
If $M$ is not very much larger than the weak scale $M_W$, we get
$g_{aUU} \sim 10^{18}\ \GeV$ for $\sum m_\nu\approx 1\ \eV$.

The axion-gluon coupling involves various three-loop diagrams, now with all six
quarks in the loop \cite{MN}. For a rough estimate we can shortcut this calculation
by integrating the effective vertex (\ref{qa1}) by parts, using the anomalous
conservation of the axial (color singlet) quark current\footnote{The actual result
  for the effective coupling (\ref{ga}) follows from a UV finite, hence non-anomalous
  3-loop diagram \cite{MN}. Within the present scheme, it is ultimately the {\em conformal
  anomaly} which accounts for the non-vanishing coupling in (\ref{ga}).}
(see e.g. \cite{Bertlmann})
\beq
\partial_\mu \left( i \bar{q}\gamma^5 \gamma^\mu q \right)
 = \frac{\alpha_s}{4\pi} \, {\rm Tr}\, G^{\mu\nu} \tG_{\mu\nu}
\; \equiv \; \cQ
\eeq
with the gluonic topological density $\cQ(x)$ (in principle there
could appear extra terms $\propto m_q \bar{q}\gamma^5 q$ on the r.h.s.,
but Goldstone's Theorem assures us that such non-derivative terms must
drop out in the final result (\ref{ga})). Summing over the six
quark flavors and now also the three leptons we thus obtain
\beq\label{ga}
\mL_{\rm eff}^{agg} =
\frac{18\a_s}{4 \pi g_a} a\, {\rm Tr}\, G^{\mu\nu} \tG_{\mu\nu}
  \equiv 18 \, g_a^{-1} a \cQ
\eeq
When the quark mass matrix $m_q$ is complex there is an extra contribution
to this term from the anomalous chiral rotation required to render the quark
mass matrix real, resulting in a shift
\beq\label{ta}
18\, g_a^{-1} a \quad \rightarrow \quad  18\, g_a^{-1} \ta \equiv
18\, g_a^{-1} a + {\rm arg} \det m_q
\eeq
Because $a$ is a Goldstone boson this shift does not affect any
other terms in the effective Lagrangian, but merely replaces $a$ by
$\ta$ in (\ref{ga}).

\section{Axion potential}

Being a Goldstone boson, the axion cannot acquire a mass in perturbation theory;
likewise its vacuum expectation value remains undetermined in perturbation theory.
However, non-perturbative effects can generate a potential for the axion and thereby
lift the vacuum degeneracy. To compute it we use the formula
\beq
\big\langle\exp F \big\rangle = \exp\left[ \langle F\rangle +
\frac12(\langle F^2\rangle - \langle F\rangle^2)\, + \, \dots\right]
\eeq
with $F\equiv18g_a^{-1}\ta\cQ$. Except for possible contributions from the weak
interactions which we ignore, there is no $G\tG$ condensate and
we have  $\langle\cQ(x)\rangle =0$ (likewise
$\langle \cQ^n \rangle$ vanishes for all odd $n$).
Hence the axion potential is
\beq\label{axpot}
V_{\text{axion}}(a) =  \frac12 m_{\text{axion}} \,\ta^2 + \cO (\ta^4)
\eeq
It is important that this potential is written as a function of the {\em shifted}
axion field $\ta$ introduced in (\ref{ta}). The axion mass is therefore
\beq
m_{\text{axion}} = 18 g_a^{-1}
\left[ \int d^4x \, \big\langle \cQ(x) \cQ(0)\big\rangle \right]^{\frac12}
\eeq
We conclude that (1) indeed $\theta\equiv\langle \ta \rangle = 0$ as required for
the solution of the strong CP problem, and (2) an axion mass term is generated
by non-perturbative effects. Although the value of the $(G\tG)^2$ condensate
is apparently not known, we can estimate
$m_{\text{axion}} \sim 18 g_a^{-1} \Lambda_{QCD}^2 \sim 10^{-8}\ \eV$,
which may be still compatible with the axion being a (cold) dark
matter candidate, at least according to standard reasoning \cite{Mu,Sikivie}, and
bearing in mind the considerable uncertainties in these numbers. From (\ref{qa2}) it
is evident that the viability of this dark matter scenario requires the Majorana
scale $M$ to be not much larger than $M_W$, in contrast to the standard see-saw
proposal \cite{Min,seesaw,Yan}. This is a main new feature of the present proposal:
if true, it could be interpreted as additional evidence for a hidden conformal
symmetry of the SM \cite{Bardeen,MN,CW}, such that the observed diversity of scales
in particle physics could be explained via quantum mechanically (or even quantum
gravitationally) induced logarithmic effects \cite{MN1}.

The main virtue of the present proposal is that it provides
a {\em single} source of explanation for axion couplings and neutrino
masses, tying together in a most economical manner features of the SM
previously thought to be unrelated. Given the known SM parameters, and
parametrizing the unknown physics in terms of just the Yukawa mass
matrices, all relevant couplings are entirely calculable in terms of
UV finite diagrams, and {\em naturally} come out to be {\em very small}
without the need for any fine tuning.

Finally, we note that all results in this Letter can be equivalently
obtained if we take the scalar field $\phi(x)$ in (\ref{phi}) to be {\em real},
absorbing the phase $a(x)$ into a redefinition of the lepton fields.
This point will be discussed in much more detail in \cite{LMN}. The
redefinition also shows that the apparent periodicity of $a(x)$ in
(\ref{phi}) is spurious because the redefined Lagrangian involves
the field $a(x)$ only through its derivatives. Rather, the periodicity
parameter for $a$ is set by the effective action (\ref{ga}) and
the fact that the gluon term is a topological density (see e.g. \cite{DiV}).

\vspace{0.2cm}
\noindent{\bf {Acknowledgments:}} AL and KAM thank the AEI for
hospitality and support during this work. We are also grateful to Pierre
Fayet for his incisive comments on a first version of this work.

\vspace{0.5cm}

\begin{center}
\includegraphics[width=12cm,viewport= 100 600 540 760,clip]{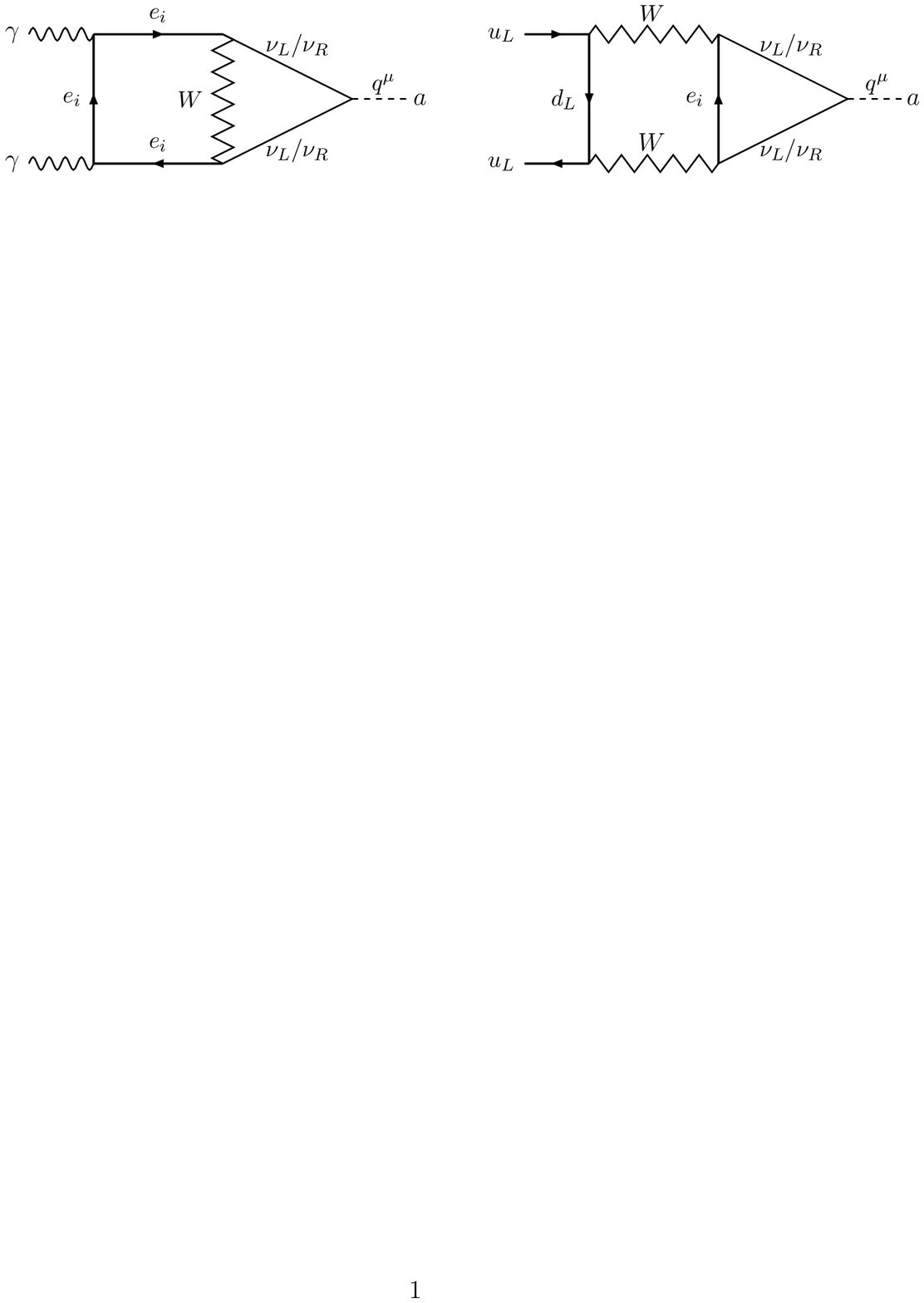}

\vspace{-1cm}
Fig.1. Axion-photon-photon and axion-quark-quark effective couplings
\end{center}

\end{document}